\documentclass{article}
\usepackage{stfloats}
\usepackage{microtype}
\usepackage{graphicx}
\usepackage{subcaption}
\usepackage{booktabs} 
\usepackage[linesnumbered,ruled,vlined]{algorithm2e} 
\usepackage{makecell}
\usepackage{hyperref}

\usepackage[preprint]{icml2026}

\usepackage{amsmath}
\usepackage{amssymb}
\usepackage{mathtools}
\usepackage{amsthm}

\usepackage[capitalize,noabbrev]{cleveref}

\theoremstyle{plain}
\newtheorem{theorem}{Theorem}[section]
\newtheorem{proposition}[theorem]{Proposition}

\theoremstyle{definition}

\theoremstyle{remark}

\usepackage[textsize=tiny]{todonotes}

\icmltitlerunning{Submission and Formatting Instructions for ICML 2026}

\begin{document}

\twocolumn[
  \icmltitle{F-scheduler: illuminating the free-lunch design space for fast sampling of diffusion models}



  \icmlsetsymbol{equal}{*}

  \begin{icmlauthorlist}
    \icmlauthor{Zilai Li}{equal,ir}
    \icmlauthor{Lujia Bai}{equal,yyy}
  \end{icmlauthorlist}

  \icmlaffiliation{yyy}{Department of Mathematics, at Ruhr University Bochum in the group of Holger Dette, Germany}
  \icmlaffiliation{ir}{Independent Researcher, Guangdong, China}

  \icmlcorrespondingauthor{Zilai Li}{lizilai2008@163.com}
  \icmlcorrespondingauthor{Lujia Bai}{lujia.bai@rub.de}

  \icmlkeywords{Machine Learning, ICML}

  \vskip 0.3in
]



\printAffiliationsAndNotice{}  

\begin{abstract}
  Diffusion models are the state-of-the-art generative models for high-resolution images, but sampling from pretrained models is computationally expensive, motivating interest in fast sampling. Although Free-U Net is a training-free enhancement for improving image quality, we find it ineffective under few-step ($<10$) sampling. We analyze the discrete diffusion ODE and propose F-scheduler, a scheduler designed for ODE solvers with Free-U Net.  Our proposed scheduler consists of a special time schedule that does not fully denoise the feature to enable the use of the KL-term in the $\beta$-VAE decoder, and the schedule of a proper inference stage for modifying the U-Net skip-connection via Free-U Net. Via information theory, we provide insights into how the better scheduled ODE solvers for the diffusion model can outperform the training-based diffusion distillation model. The newly proposed scheduler is compatible with most of the few-step ODE solvers and can sample a 1024 x 1024-resolution image in 6 steps and a 512 x 512-resolution image in 5 steps when it applies to DPM++ 2m and UniPC, with an FID result that outperforms the SOTA distillation models and the 20-step DPM++ 2m solver, respectively.  
\end{abstract}

\section{Introduction}
The diffusion probability model (DPM) proposed by \citet{2020ddpm} is a state-of-the-art image generation model.
However, compared to typical algorithms \cite{goodfellow2020gan, kingma2013auto}, sampling from DPMs requires multiple neural network evaluations and incurs substantial computational cost. Hence, accelerating the sampling from DPMs remains an active research area. 

Research related to it can be categorized into two classes. The first class proposes a training-free algorithm aimed at providing a good SDE and ODE solver tailored to the diffusion process (\citet{2020ddpm,lu2025dpm,song2020ddim,zhao2023unipc,karras2022elucidating,liu2022pseudo}). The second class deems the whole inference process
as a large neural network, and reduces the inference steps by training a small distillation model (\citet{luo2311lcm,song2023improvedlcm,lin2024sdxllightning,yin2024dmd,yin2024improveddmd,salimans2022progressive,sauer2024adversarial,kim2023consistency,song2023consistency,liu2023instaflow}), which can generate images in four to eight steps. 

Most training-free ODE solvers are not tailored to diffusion models, except the method of \citet{karras2022elucidating}, which shows that a uniform discretizing schedule causes increasing truncation error in few-step inference and proposes a non-uniform schedule with larger steps in high-noise regions and smaller steps in low-noise regions.
However, this method is not customized for the latent diffusion model.  

The latent diffusion model itself includes $\beta$-VAE and skip connections, both of which can be additionally boosted by a better scheduler in the design space.    
The $\beta$-VAE trained with a little noise can also reduce the inference budget by preventing the total denoising of images.  However, adding same noise to the latent feature at different resolutions also has distinct effects\cite{hoogeboom2023simple}, which necessitates the investigation of the design space. Furthermore, we find that the U-Net decorator, which aims to change the skip connection to enhance the denoising process, has the side effect of mimicking the distillation model\cite{ma2024surprising} and changing the discretization result.  We use ODE theory to fix the unwilling discretization effect, and use information theory to explain why the distillation model has the diversity problem, which provides an explanation of the phenomenon proposed by \citet{bertrand2025closed}, and proves the advantage of exploiting the design space of using a skip-connection decorator.

\begin{figure}[ht]
  \vskip 0.2in
  \begin{center}
    \centerline{\includegraphics[width=\columnwidth]{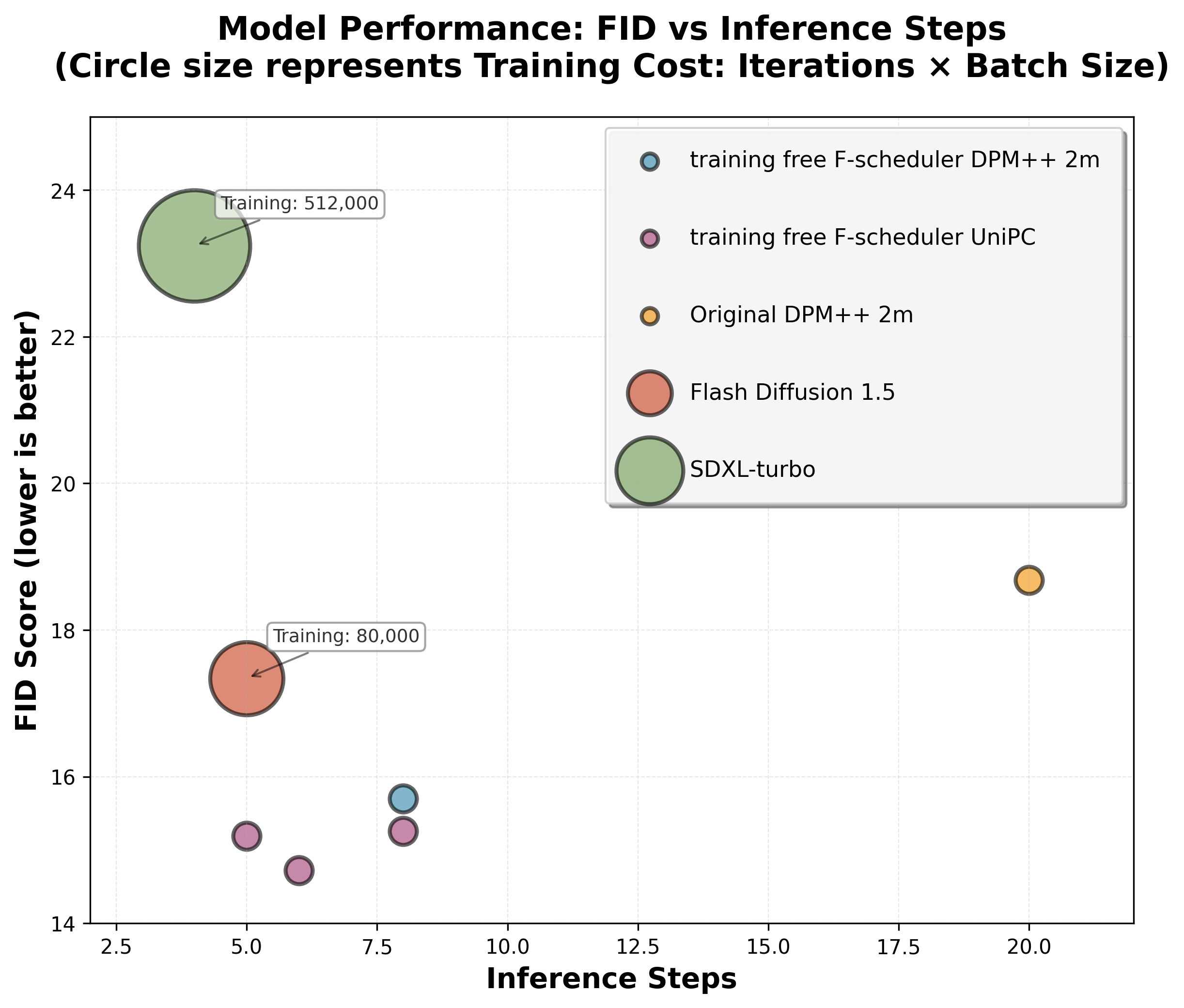}}
  \caption{The comparison of FID performance using Free-U in different steps via ODE solvers in COCO 2014 with guidance scale 5.5.  The size of the circle is proportional tothe amount of training cost.}
  \label{compareFIDResult}
  \end{center}
\end{figure}

To that end, by exploring the design space for the latent diffusion model, we implement a new inference method compatible with most ODE solvers. Applied to UniPC\cite{zhao2023unipc}, it can generate 512 x 512 images in 5 steps without additional training, and its Frechet Inception Distance (FID) (\citet{heusel2017fid}) performance surpasses both the DPM++ 2m solver and the SOTA distillation model in 20 steps and 8 steps, respectively, see Fig \ref{compareFIDResult}. 

\begin{figure}[ht]
  \vskip 0.2in
  \begin{center}
    \centerline{\includegraphics[width=\columnwidth]{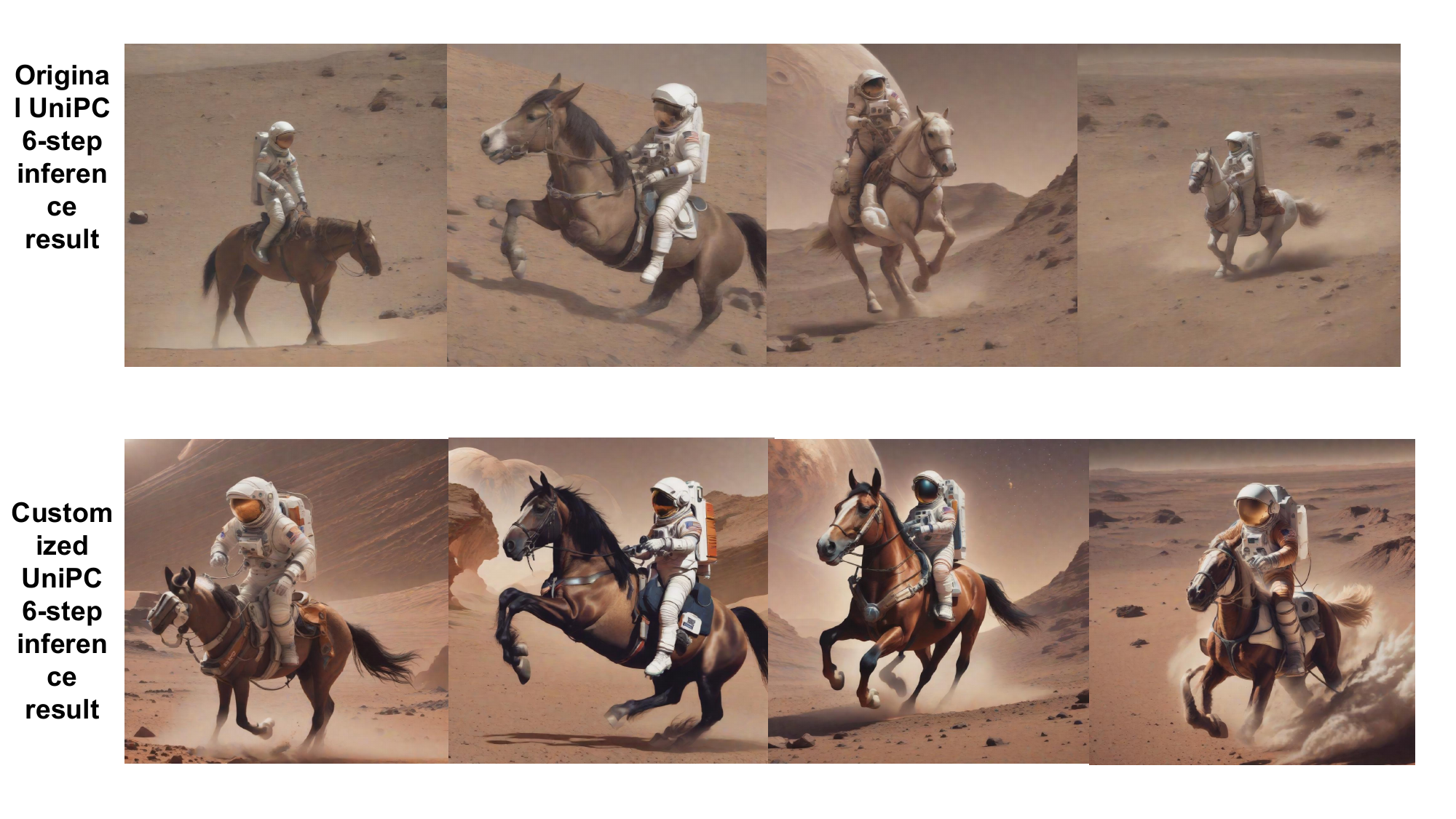}}
  \caption{The comparison between our algorithm and UniPC solver in few step inference with the prompt "an astronaut riding a horse on Mars".}
  \label{compareResult}
  \end{center}
\end{figure}

We also apply our algorithm to 1024 x 1024 image generation, and the FID performance of the eight-step and six-step inference exceeds most of the latest distillation models. We perform extensive experiments using the COCO 2014 (\citet{lin2014coco}), COCO 2017 (\citet{lin2014coco}), and LAION datasets (\citet{schuhmann2022laion}), and exhibit the generation results in Fig \ref{compareResult}. 

\section{Related Work}

\subsection{Diffusion ODE and ODE Solver}
The image-generation process of the diffusion model can be deemed as solving a special diffusion ODE. A typical diffusion model has two processes: forward and reverse. The forward process perturbs the image by adding Gaussian noise to it.  Hence, in the timestep $t$, the probability density of the noisy feature $x_t$ with given $x_0$ belongs to a Gaussian distribution described in the \cref{eq:probabilityDefine}.  
\begin{equation}
        q(x_t|x_0) = \mathcal{N}(\alpha_tx_0;\sigma^2_tI).
        \label{eq:probabilityDefine}
\end{equation}
where $\alpha_t$ and $\sigma_t$ can be choose arbitrarily if it can ensure the final $p(x_T|x_0)$ close to a standard Gaussian distribution.

In contrast, the reverse process, which corresponds to generating an image from pure Gaussian noise, can be described by a time-dependent ODE. Via the research from \citet{lu2022dpm}, when given a noisy input $x_s$ and a target timestep $t$, the correct noisy feature output $x_t$ can be gained by solving the \cref{eq:common_x} :
\begin{equation}
   x_{t} = \frac{\alpha({t})}{\alpha(s)}x_s - \alpha_t\int^{\lambda_t}_{\lambda_s}e^{-\lambda} \epsilon_\theta(x_{t(\lambda)},\lambda)d\lambda,
  \label{eq:common_x}
\end{equation}

where $\lambda_t$ is the log Signal-to-Noise Ratio (SNR) of the noisy feature $x_t$,and the $\epsilon_\theta$ is modeled by a neural network with ground truth proportional to the score function $\nabla_x \log{q(x_t)}$, which is trained with the objective function described in \cref{eq:scoreFunctionDefine} 
\begin{equation}
   \mathcal{L}_\theta = E_{x_0}[q(x_t|x_0)||\frac{\epsilon_\theta(x_{t},t)}{\sigma_t}-\nabla_x\log q(x_t|x_0)||^2],
  \label{eq:scoreFunctionDefine}
\end{equation}
By using the Taylor Expand in the $\epsilon_\theta$ (\citet{lu2022dpm,lu2025dpm}), we get the following \cref{eq:taylor_version_xt}:

\begin{align}
    x_{t} &= \frac{\alpha({t})}{\alpha(s)}x_s - \alpha_t \Sigma_{n=0}^{k-1}{\epsilon}^{(n)}_\theta(x_{t(\lambda)},\lambda) \int^{\lambda_t}_{\lambda_s}e^{-\lambda} \frac{(\lambda-\lambda_s)^n}{n!}d\lambda \nonumber
    \\& + \mathcal{O}(h_i^{k+1}),
  \label{eq:taylor_version_xt}
\end{align} where $h_i \propto \lambda_t-\lambda_s$

However, calculating the neural networks' deviation requires substantial computational resources.  Researchers typically use different algorithm, like the Runge-Kutta method (\citet{lu2025dpm, zhao2023unipc}), to approximate the deviation.



\subsection{Time Schedule}
The original diffusion ODE describes a continuous curved trajectory, and the \cref{eq:common_x} approximates it by discretizing the path with a series of lines. Research on reverse ODE solvers aims to precisely estimate the next step, thereby minimizing truncation error from pool discretization in the few-step inference.  On the other hand, the discrete method proposed by \citet{karras2022elucidating} does not aim at design a new ODE solving algorithm, but proposes a choice of moving distance $h_t$ at each step $t$ to enhance the sample quality.  Via the analysis of the Karras' research, the $h_t$ should be large at the beginning of the inference, and then decrease monotonically.  Since truncation error is bound by the \cref{eq:errorLimit}
\begin{equation}
  ||e_N||\leq E max_i||\mathcal{T}_i||
  \label{eq:errorLimit},
\end{equation}
where $||\mathcal{T}_i||\leq Ch_i$.  The $E$ depends on the discrete step $N$, the start time step $t_0$, the ending timestep $t_N$, and the existing constant $C$.  The meaning of this upper bound is limited by the part that contributes the highest $\mathcal{T}_i$.  And \citet{karras2022elucidating} reveals that the truncation error of the diffusion model increases monotonously when the $h_i$ in each inference step is equal.
Based on this observation, \citet{karras2022elucidating} proposes a following discrete method:
\begin{equation}
   \sigma_i = (\sigma_{min}^{\frac{1}{p}}+\frac{i}{N-1}(\sigma_{min}^{\frac{1}{p}}-\sigma_{max}^{\frac{1}{p}}))^p.
  \label{eq:karrasSchedule}
\end{equation}
Moreover, according to the observation from the \citet{li2025direct}, if we choose $p=7$ and $N=20$, the fourth-to-last step corresponds to the first $\frac{6}{1000}$ perturbation steps, and it will cost $\frac{1}{5}$ inference budget. 
Based on that observation, \citet{li2025direct} proposes another discrete method:
\begin{equation}
   t_i = (t_{min}^{\frac{1}{p}}+\frac{i}{N-1}(t_{min}^{\frac{1}{p}}-t_{max}^{\frac{1}{p}}))^p.
  \label{eq:improvedKarrasSchedule}
\end{equation}
When choosing $p=1.2$, the $h_i$ in both the noisy part and the less noisy part are larger than those in Karras's original method. Hence, it can address the problem with the original Karras schedule mentioned above.

\subsection{Diffusion Distillation}
The training-free method optimizes the truncation error at each step and thus minimizes the need for additional inference steps for the precise discretization. Unlike this method, the distillation algorithm (\citet{lin2024sdxllightning,sauer2024adversarial,luo2311lcm,song2023improvedlcm,salimans2022progressive}) considers the whole inference process of applying a neural network iteratively as a huge neural network. The distillation algorithm, which typically trains a small neural network, reduces its inference steps.  However, training a neural network with high-quality content, fast inference, and no additional parameters is difficult. Actually, the distillation algorithm, through the analysis of information theory (\citet{li2025direct}) and experiments, can sometimes reduce the diversity of the generated images, and affect the FID performance. We do further analysis and experiments to enhance the augmentation.

\subsection{Latent Diffusion Model}
Another method for reducing the computational requirements of inference is to encode images into a low-dimensional latent space.  Stable diffusion performs inference in a 64 x 64 x 4 latent space to generate a 512 x 512 x 4 image.  To achieve this, it firstly encodes the image into a low-dimensional latent space via a $\beta$-variational autoencoder (\citet {burgess2018understandingBeta}) to let the following inference ignore the perceptual detail. Then, it performs diffusion inference in latent space via a U-Net (\citet{ronneberger2015unet}) with an attention mechanism to model the semantic information. Experiments show that compared with vector quantify VAE (\citet{2020vqvae}), the normal $\beta$-VAE with a small KL-term constraint generates higher-quality images.
\section{Approach}
\subsection{The effect of the skip-connection modification} 
\begin{figure}[ht]
  \vskip 0.2in
  \begin{center}
    \centerline{\includegraphics[width=\columnwidth]{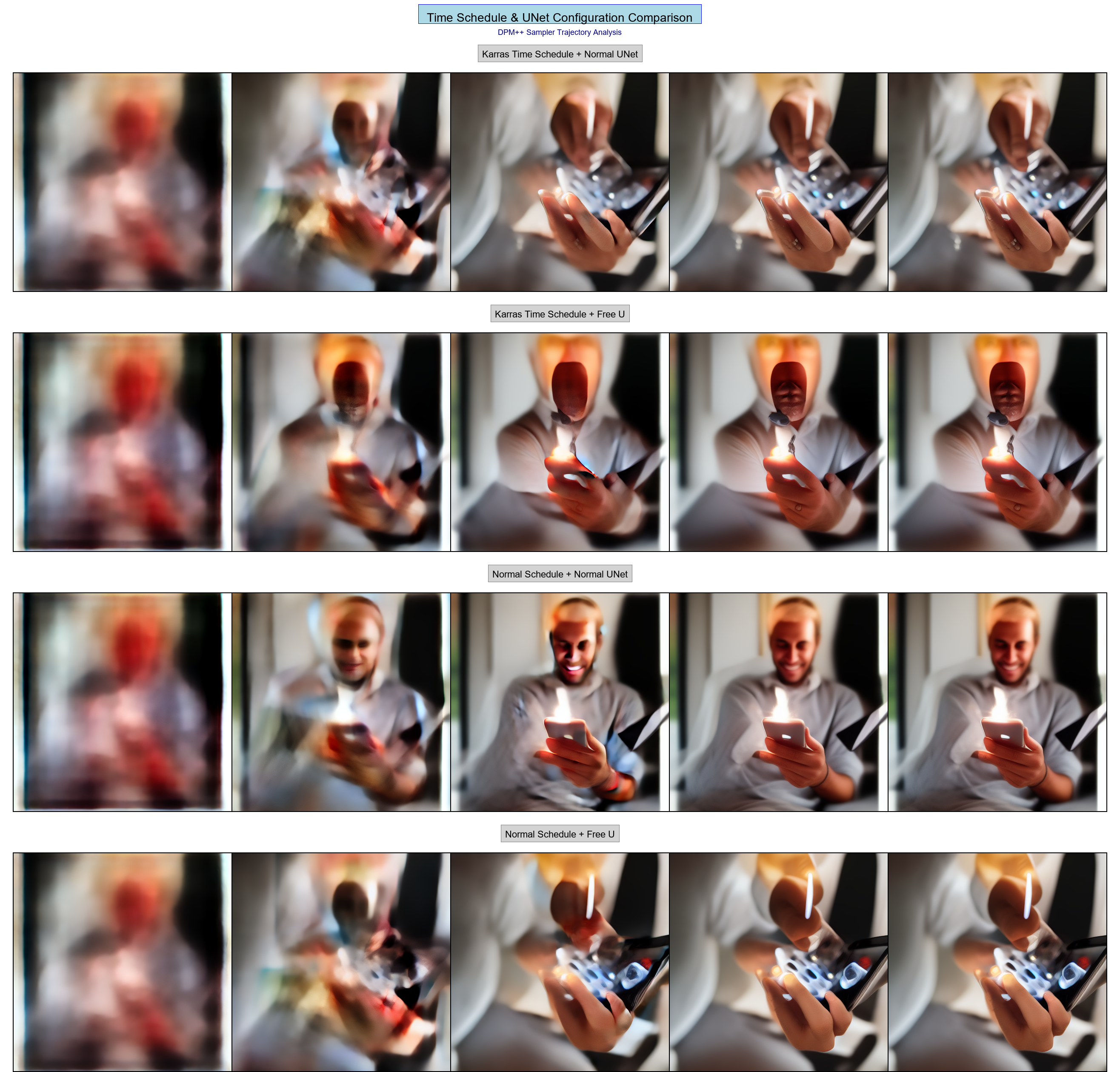}}
  \caption{The comparison between different discrete methods when using the same DPM++ 1s Sampler in a few-step inference.  The first line exhibits the trajectory of the normal Karras discrete method.  The second one exhibits the trajectory of the Karras discrete method + Free-U, which has an unwilling result. The third exhibits the normal discrete method with a normal U-Net, and the final one exhibits the normal discrete method with Free-U.  And the final line is similar to the first line.}
  \label{compareDifferentTimeSchedule}
  \end{center}
\end{figure}

Modifying the skip connection via Free-U\cite{si2024freeu}, a training-free U-Net decorator, can reduce the low-frequency component of the noisy skip connection and enhance the backbone feature, thereby augmenting the denoising capability of the diffusion model while preserving image details\cite{si2024freeu}.  We find that the skip connection modification removes the blur from the generative image, implying a change in the neural network's original output from a standard score function to a vector pointing to a special image, which allows the model's behaviour to be similar to that of the diffusion disillation model. 
However, experiments in \cref{compareDifferentTimeSchedule} show that Free-U will also change the discrete method of the diffusion ODE, and this effect cannot be ignored in the few-step inference that has a poor discrete situation.


Explicitly, the output of the neural network in \cref{eq:scoreFunctionDefine} is pointing to the average of all possible images that can cause this noisy image.  In an extreme few-step inference case, like one or two NFE inference, the final output will be blurred, since in that step, the score function only points to the centroid of images.  Meanwhile, the distillation algorithm, like rectify flow proposed by \citet{liu2022rectified}, will tend to modify the output and let it point to a special sample.

Although the task of modifying the score function to allow it to point to a special sample without additional training is hard, the Free-U that preserves the detail of the output image while augmenting the denoise capability can partly achieve it in the least noisy stage, since the possible images are very limited.  The previous research from \citet{ma2024surprising} also shows that compared to the original diffusion model, the diffusion distillation model has small skip connection scalers.

\begin{figure}[ht]
  \vskip 0.2in
  \begin{center}
    \centerline{\includegraphics[width=\columnwidth]{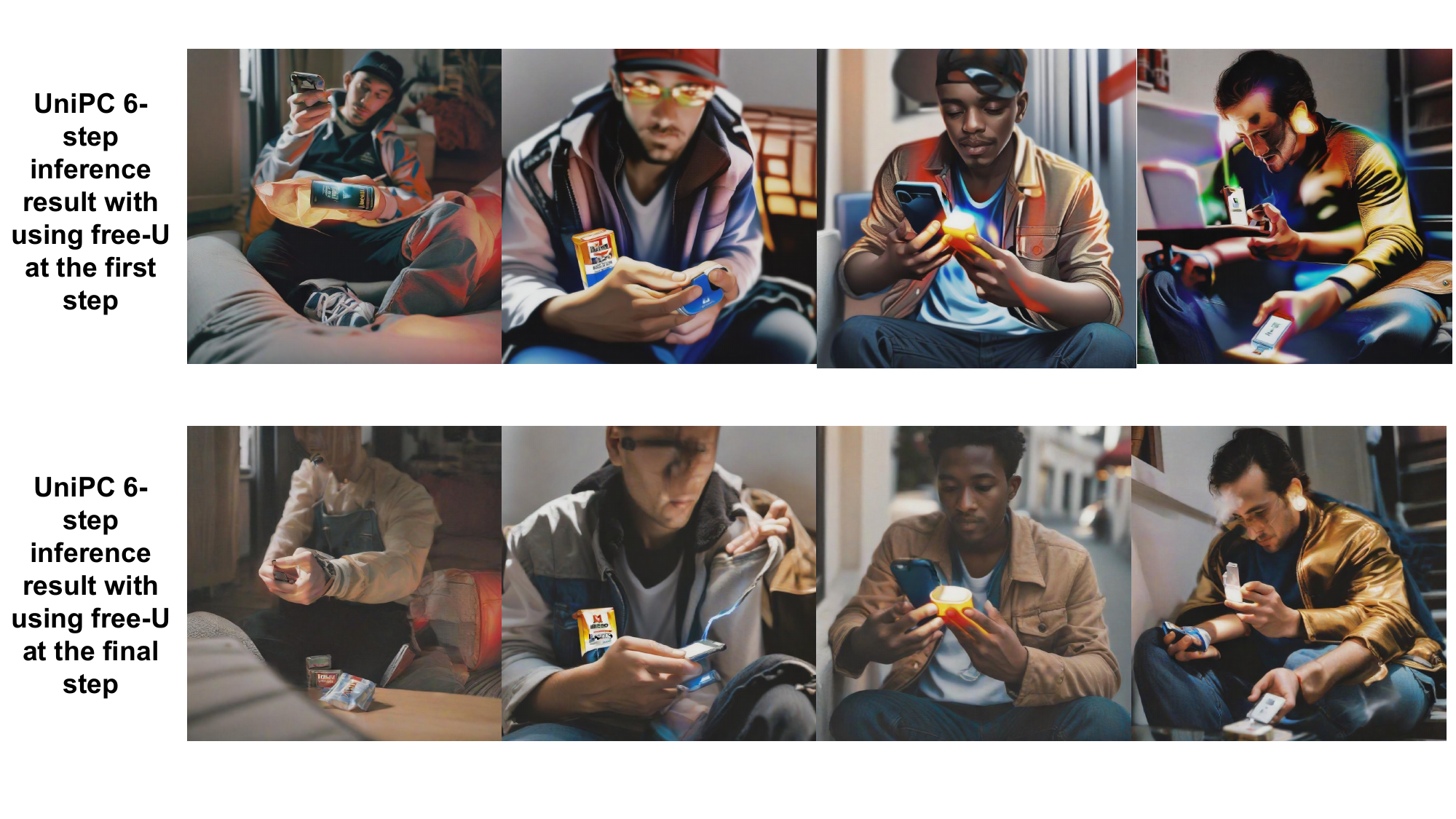}}
  \caption{The result that starts using free-U in the different steps of the UniPC solver.}
  \label{compareDifferentStepInUniPC}
  \end{center}
\end{figure}

However, we don't know which step could be changed simply from the score function to the conditional score function pointing to special a sample, and one of the vanilla ways is to apply free-U at all time steps, as the original paper claims. This method has been seen as helpful to slightly modify the score function and make it more similar to the distillation's output.
But, via the mathematical analysis of the diffusion ODE and the experimental result in \cref{compareDifferentStepInUniPC}.  This vanilla method in the extreme few-step inference of high-resolution images, such as 6 NFE, would not work properly.

Explicitly, consider that the normal ResNet (\citet{he2016res}) can be viewed as an ODE (\citet{chen2018neuralODE}) by using the following formula:
\begin{equation}
   l_{t+1} = l_{t} + f_\theta(l_t,t)
  \label{eq:normalRes},
\end{equation}
\begin{equation}
   \frac{dl}{dt} = F_\theta(l_t,t),
  \label{eq:neuralODE}
\end{equation}
where $l_t$ is the feature output in layer $t$.

The U-Net, as a black box ODE solver using skip connections, can be regarded as another ODE. And modifying the skip connection and scale backbone feature in the neural ODE will further increase the $h_i$ in \cref{eq:taylor_version_xt}. 

To further show the influence of the Free-U on the black box ODE solver, we then visualize the decoded trajectory using different discrete methods by utilizing the same sampler DPM++ 1s sampler in \cref{compareDifferentTimeSchedule}. The uniform time schedule using Free-U generates a feature in the first two steps that is similar to the \citet{karras2022elucidating} time schedule without Free-U, while totally different to the normal uniform sampler without using Free-U, providing evidence that Free-U increases the moving distance at the beginning of the inference. And if we use karras or a similar time schedule, it will have an unwilling effect, since it's large enough at the beginning step.

To prevent this problem, another method is to parameterize the start step of using the U-Net decorator in an existing solver.  However, experimental results show that this Vallina method doesn't work either.  We visualize the 6-step inference result combining the free-U and UniPC solver in \cref{compareDifferentStepInUniPC}.

To address black-box ODE-solving problems, we should provide a new time schedule. 

\subsection{The time schedule leveraging $\beta$-VAE}
\begin{figure}[ht]
  \vskip 0.2in
  \begin{center}
    \centerline{\includegraphics[width=\columnwidth]{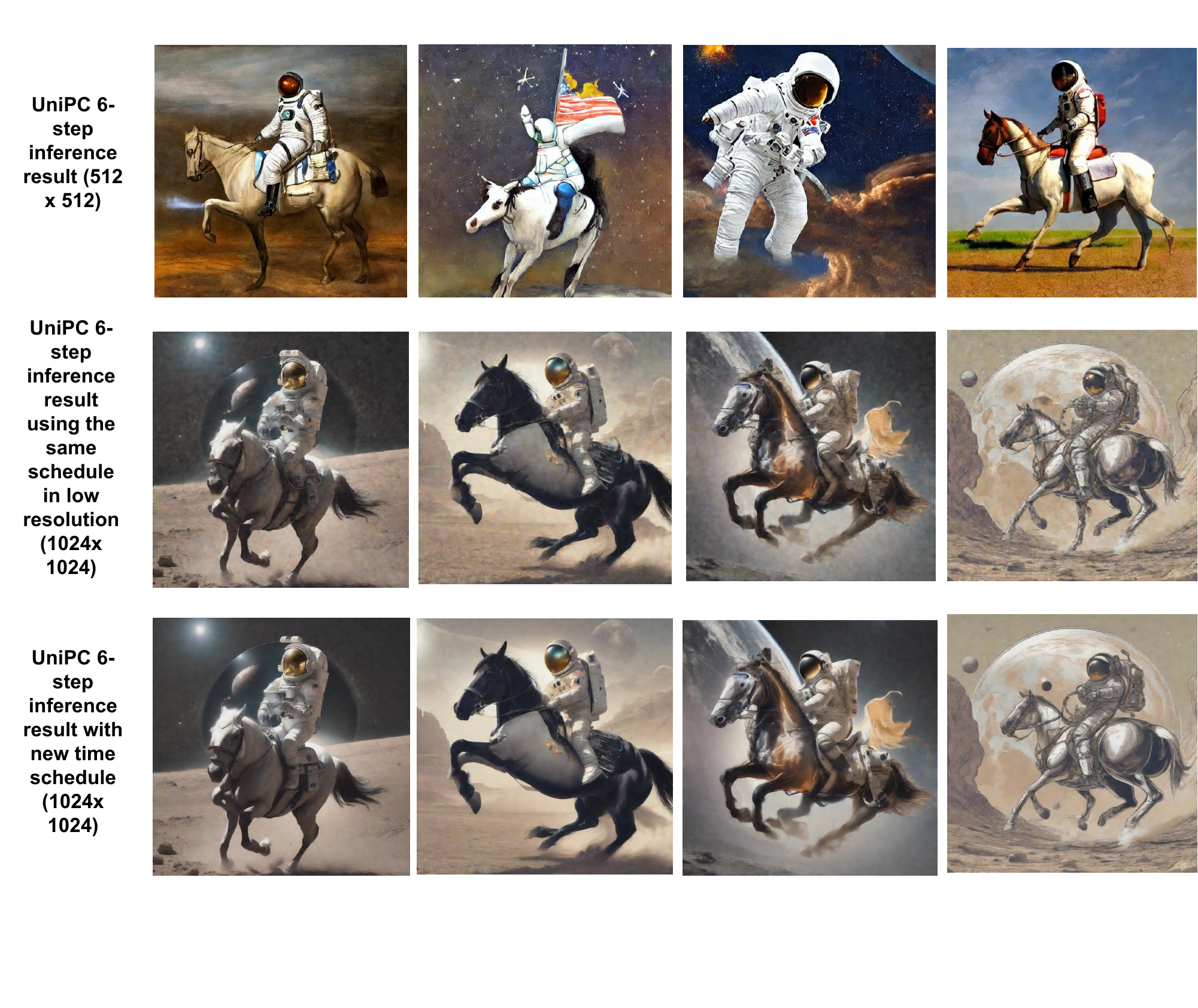}}
  \caption{The comparison between the different resolution images generated by the same time schedule and ODE solver utilizing the same $\beta$-VAE.}
  \label{compareResultInVae}
  \end{center}
\end{figure}

\begin{figure}[ht]
  \vskip 0.2in
  \begin{center}
    \centerline{\includegraphics[width=\columnwidth]{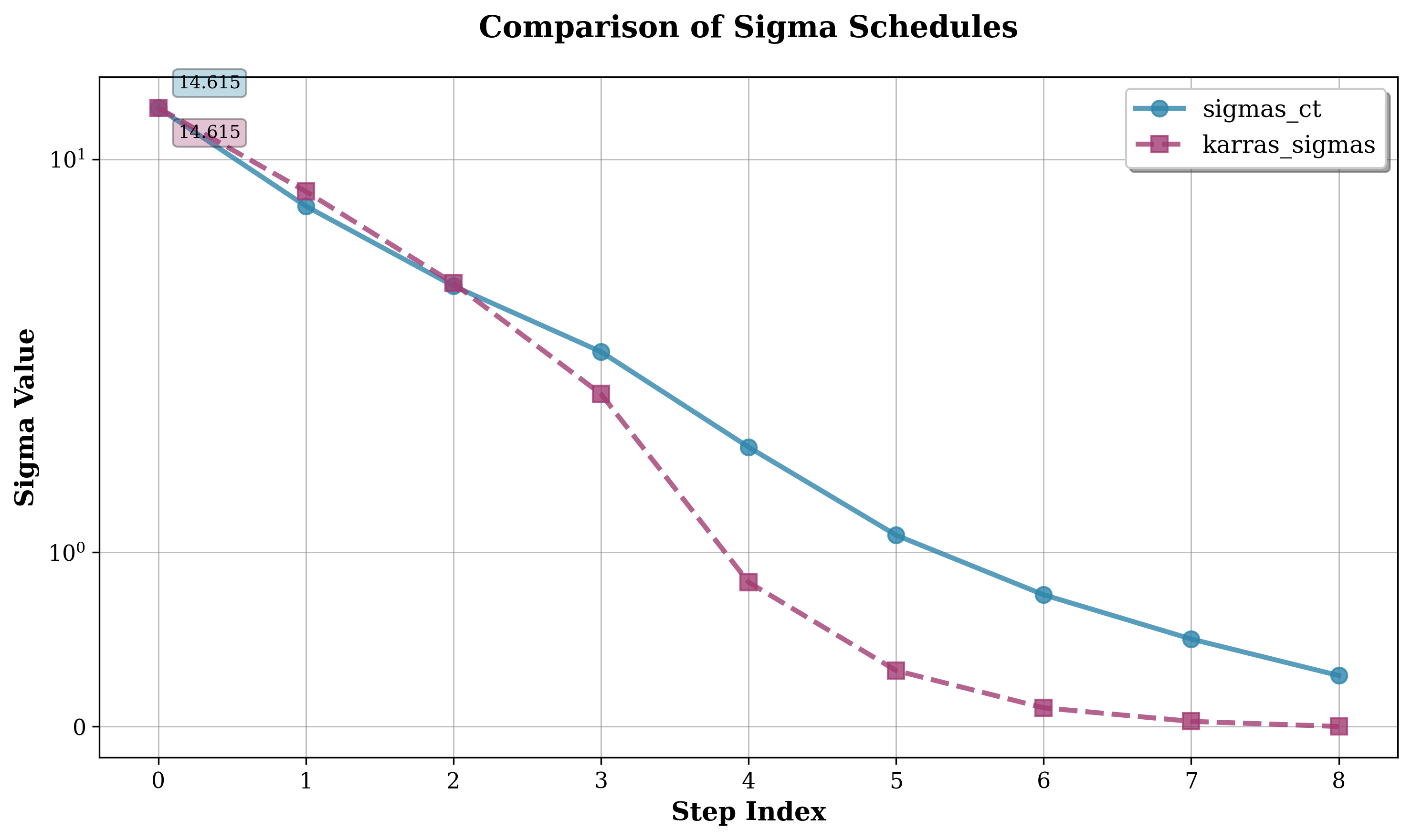}}
  \caption{The comparison between our time schedule and the time schedule proposed by Karras}
  \label{compareSigma}
  \end{center}
\end{figure}

The decoder of the latent diffusion model, a $\beta$-VAE trained with a small KL-term constraint, can handle a small amount of noise; thereby, if we choose a time schedule that does not fully denoise the image, we can save further inference budget. Although the latent diffusion model theoretically does not require using the KL term to train an autoencoder, this optimization trick is very prevalent in the training of the diffusion model.  The latest representation autoencoders (RAE) also add noise during the training to augment the decoder \cite{zheng2025rae}.  The experimental results show that the decoder's capacity to handle noisy features varies with image resolution.  In the \cref{compareResultInVae}, the second line uses the same time schedule as the first line, which doesn't denoise totally, but those images in the second line have some blurred parts that harm the details of the images, while the first line doesn't confront this problem.  And the third line output feature has less noise and produces better results. 
Based on the aforementioned idea and the experiment results in Section 4, we first involve the following discrete method that aims to correct the output in the less noisy stage:
\begin{equation}
   t_i = (t_{max}^{\frac{1}{p_1}}+\frac{i}{N}(t(\sigma_{stop})^{\frac{1}{p_1}}-t_{max}^{\frac{1}{p_1}}))^{p_1},
  \label{eq:finalDiscrete}
\end{equation}
where $\sigma_{stop}$ is defined via \cref{eq:sigmaDefine}, and $t(\sigma)$ is a function projecting variant $\sigma$ to timestep $t$, which use the discrete version describe in \cref{eq:probabilityDefine} whose range is between $[1,1000]$:
\begin{equation}
   \sigma_{stop} = (\sigma_{max}^{\frac{1}{p_2}}+\frac{stop}{M}(\sigma_{min}^{\frac{1}{p_2}}-\sigma_{max}^{\frac{1}{p_2}}))^{p_2},
  \label{eq:sigmaDefine}
\end{equation}
and $\sigma$ could be nonzero to save the one-step inference budget by utilizing $\beta$-VAE.  Meanwhile, the problems of using Free-U while preventing an increase in moving distance could be addressed by applying this special time schedule, since we could only apply Free-U at timesteps that move a short distance.  By leveraging the robustness of the $\beta$-VAE, the total moving distance has also been constrained.
The visual difference between our discrete method applying on DPM++ 2m, and the method proposed by the Karras' paper is in \cref{compareSigma}
Meanwhile, since the total moving distance should be as limited as possible, we also apply the analytical first step (AFS) proposed by \citet{dockhorn2022genie} when generating high-resolution images.

Our algorithm can be described by the \cref{tab:alorithm}
\begin{algorithm}
  \SetAlgoLined
  \KwData{number of inference step $N$, initiated noise $\epsilon$, stop step in Karras schedule $t_{stop}$, the total step $M$ of the original discrete method, discretion hyper parameter $p_1$ and $p_2$, U-Net augment step $t_{aug}$, ODE solver $F_\theta$, augment U-Net method $freeU()$, a bool value $afs$ to determine whether using afs, and an analytical first step solver without using neural network $F$}
  \KwResult{final result image }
  \LinesNumbered
   $\sigma_{stop} = (\sigma_{max}^{\frac{1}{p_2}}+\frac{t_{stop}}{N+t_{stop}+1}(\sigma_{min}^{\frac{1}{p_2}}-\sigma_{max}^{\frac{1}{p_2}}))^{p_2}$\;
   
  $
   ts =\{t_i = (t_{max}^{\frac{1}{p_1}}+\frac{i}{N}(t(\sigma_{stop})^{\frac{1}{p_1}}-t_{max}^{\frac{1}{p_1}}))^{p_1};0\leq i < N \}$

   \If {$afs$} {
        $ts.insert((\frac{t_0+t_1}{2}),at: 1) $
   }
   
  $x = \epsilon$
  
  \While{$t$ in $ts$}{
      \lIf{$t$ equal to $t_{aug}$}{
        $freeU(F_\theta)$\;
      }
      \lIf{$afs$ and $t$ equal to $999$} {
      $x = F(x)$
      }
      \lElse {
      $x = F_\theta(x)$ 
      }
    }
  \caption{F-scheduler in the common framework of Diffusion ODE solving}
  \label{tab:alorithm}
\end{algorithm}
\subsection{Information Theory Analysis of the Different Algorithms}
The experimental results in the \cref{sec:Experiments} show that the diversity of the generative images sampled from a distillation model is the main bottleneck of its FID performance. To explain why our algorithm can outperform the distillation model, we use information theory to provide a tentative analysis of the diffusion model.

\begin{figure}[ht]
  \vskip 0.2in
  \begin{center}
    \centerline{\includegraphics[width=\columnwidth]{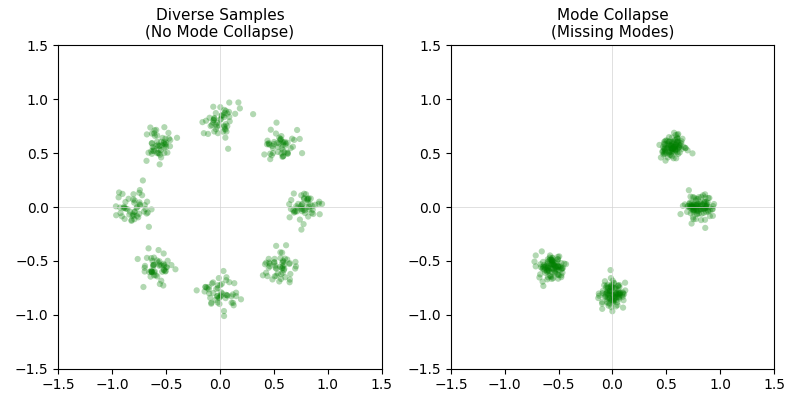}}
  \caption{Illustration of Mode Collapse. (Left) The target distribution consists of a mixture of 8 Gaussians arranged in a ring. (Right) A simulated example of mode collapse that contains only a subset of the available modes.}
  \label{modeCollapse}
  \end{center}
\end{figure}

\citet{bertrand2025closed} shows that using the ground truth score function in the place having the highest loss, explicitly, 20\% of inference, to replace the studied score function will harm the diversity of the diffusion model.  They don't provide a theoretical analysis of this phenomenon, nor do they provide a further estimate of the distillation model that modifies the score function. 
In this section, the information bottleneck theory  \cref{thm:loss} reveals that mutual information affects the model's generalization.  
We find out when the score function been changed from the centroid of all possible images that can generate the current noisy image to a special image, the mutual information between the middle feature and the output increases, as described in Proposition \ref{eq:lVIB}.  
By preventing the mutual information from affecting, \cref{prop:loss} and Proposition \ref{eq:decomposeLose} show that the current average output of the model would not shift and deviate from the ground truth centroid,  while the mode collapse in GAN, shown in \cref{modeCollapse}, is different from that.
\citet{jeon2021ib} also shows that constraining the information bottleneck in GAN can enhance the FID. 

Eplicilty, for an arbitrary image generative model, we assume that there is a ground truth noise $x$ and image $y$ pair that is possible to be implicitly studied by the neural network, and $y$ sample from $\mathcal{Y}$ contains images with enough diversity.

Then, according to learning theory, if we want to prevent a model from rote, we should constrain the number of functions this model can express, to ensure it can get access to the best hypothesis in training. 

Suppose we have a training data $s_n=\{y_i\}_{i=1}^n$ that has at least $1-\sigma$ probability drawn from the same distribution of $\mathcal{Y}$, a neural network $f^s$ trained in $s$ via SGD in $i$ iteration, a batch size $m$ and a batch validating data $s_m=\{\hat{y}_i\}_{i=1}^m$, a middle feature $Z_l^s$ output from the layer $l$ of the neural network $f^s$. Then, for the output of the $f^s$, we consider a output $[f^s(\hat{x}_1),f^s(\hat{x}_2),...,f^s(\hat{x}_m)]$ from a network on $i$-th iteration traning which the $m$ noise-image pair $\hat{x},\hat{y}$ is denote by the ground truth matching. We consider the finite hypothesis set $\mathcal{H}$ that consists of all possible outputs in the $i$-th iteration of SGD training, without considering the order of outputs within the same batch.  Meanwhile, in SGD, the training process is random, and with a given noise-feature pair $x, z_l^s$ in training, we assume $p(Z_l^s|Y)$, who denote the probability density of $Z_l^s$ in the $i$ iteration of SGD training, will use $z_l^s$ as the means.  And the mutual information $I(Y; Z_l^s)$ is given by $p(Z_l^s, Y)$. Finally, seperate the neural network into two parts: decoder $g^s$ and encoder $\phi^s$, and $f^s=g^s \circ \phi^s$, where we use $\circ$ to denote the composition of functions.  

First, the following theorem investigates the information bottleneck (IB) of the latent diffusion model, whose proof is listed in the Appendix A.
\begin{theorem} \label{thm:loss}
For the L$_2$ loss function $\mathcal{L}$, and $\Delta(s)=E_{X, Y}[\mathcal{L}(f^s(X) ,Y)] - \frac{1}{m} \Sigma_{i=1}^m\mathcal{L}(f^s(x_i),y_i)$,  we have:
    \begin{equation}
   \Delta(s)\leq \underset{l}{\arg\min} \sqrt{\frac{2^{mI(Y;Z_l^s)}+{I(\phi_l^s; s_n)}+log\frac{2}{\sigma}}{2o* A_m^m}},
  \label{eq:RealIBBoundFull}
\end{equation}
where $o$ is a fixed constant for every hypothesis space, and $A_m^m$ denotes the  number of arrangements of $m$ distinct objects. 
\label{prop:loss}
\end{theorem}

 Unlike \cite{kawaguchi2023does} that applies IB to a stochastic neural network, we also show that, even without the stochastic part in the neural network, the SGD itself will give the mutual information between the middle feature and the input, and it still constrains the possible hypotheses.  Add a stochastic process to the neural network, or use the loss with high variance, just a special case of our analysis.  
 
Then, it is intuitive that changing the direction of the score function to allow it to point to the final result, rather than the centroid of the possible image, will affect the mutual information, then affect the \cref{eq:RealIBBound}, and cause the diversity problem. We provide further analysis regarding this.
Explicitly, the diffusion ODE can be regarded as a Markov process $Z_\theta(T) - Z_\theta(T-1)-...-Z_\theta(t)-...Z_\theta(0)$, in which $Z_\theta(t)$ is the noisy input of the diffusion model, and we have the following Proposition \ref{eq:lVIB}:
\begin{proposition}
$I(Z_\theta(0),Z_\theta(t)) = H(Z_\theta(0)) - H(Z_\theta(0)|Z_\theta(t)) \leq H(Z(0)) - E_{Z_\theta(t)}[-p(Z_\theta(0)|Z_\theta(t)) * \log p_\theta(Z_\theta(0)|Z_\theta(t)))]$
  \label{eq:lVIB}
\end{proposition}
in which $H(Z(0))$ is the entropy of the real $Z(0)$ sample from the dataset. If we modify the $Z_\theta(t)$ to make it similar to the final output to enhance the log-likelihood by rectifying the trajectory, then the upper bound of the mutual information will also increase.

In the following proposition, we decompose the $E_{X,Y}[\mathcal{L}]=E_{X, Y}[\mathcal{L}(f^s(X) ,Y)]$ into two parts to explain mode collapse.  And $\Delta(s)$ been constrain by IB above is the uppper bound of the $E_{X,Y}[\mathcal{L}]$
\begin{proposition}\label{prop:collapse}
$E_X[(f^s(X)-\hat{f}(X))^2]=E_{X,Y}[\mathcal{L}] - E_{X,Y}[(\hat{f}(X)-y)^2]$, where $\hat{f}(x)=E_{s_n}[f^s(x)]$
  \label{eq:decomposeLose}
\end{proposition}
Meanwhile, as the Proposition \ref{eq:lVIB} explain, the constrain of mutual information also reduce the log likelihood related to the second term.  And, in GAN, the mode collapse will enlarge the $E_X[(f^s(X)-\hat{f}(X))^2]$, like the \cref{modeCollapse}.  Hence Proposition \ref{eq:decomposeLose} can prevent the diversity problem. The proof of Proposition \ref{eq:decomposeLose} is postponed to Appendix B.

\section{Experiments}
\label{sec:Experiments}

\subsection{Ablation Study}
To demonstrate the necessity of applying Free-U at different steps, we conduct an ablation Study on a customized UniPC solver that aims to generate a 1024 x 1024 image in a few-step inference guided by 5000 prompts sampled from the COCO 2017 dataset.  Explicitly, for the 8-step inference, we employ $p_1=12$, $p_2=1.2$, $stop = 10$, $M=12$, and $N=8$ as the discrete setting.  And for the 6-step inference, the discrete setting is $p_1=12$, $p_2=1.2$, $ stop=10$, $M=12$, and $N=6$.  Then, we try the following setting: 1. apply Free-U on every step of inference, 2. apply Free-U only on the last few steps of inference, 3. Don't apply Free-U. The experimental result is in the \cref{tab:ablation}.  The $w$ in the \cref{tab:ablation} is the guidance scale of inference, and the number in the parentheses is the total number of function evaluations (NFEs) of the model.  Meanwhile, for the 8-step UniPC solver, we use a 2nd-order multistep algorithm, whereas the 6-step solver is 1st-order solver.
\begin{table}[htbp]
  \caption{Results of applying Free-U in different steps of inference.}
  \label{tab:ablation}
  \begin{center}
  \resizebox{\columnwidth}{!}{
  \begin{tabular}{@{}lccccc@{}}
    \toprule
    Solver & steps & use Free-U & apply stage & FID & CLIP-Score \\
    \midrule
    UniPC & 6 & Yes & Every Steps & 28.88 & 30.96\\
    UniPC & 8 & Yes & Every Steps & 33.70 & 30.5\\
    UniPC & 6 & No & Never  & 27.13 &  31.49\\
    UniPC & 8 & No & Never  & 25.47 &  31.39 \\
    UniPC & 6 & Yes & 4th step  & 25.20 & 31.42\\
    \textbf{UniPC} & \textbf{8} & \textbf{Yes} & \textbf{6th step} & \textbf{24.38} & \textbf{31.38} \\
    \bottomrule
  \end{tabular}
  }
  \end{center}
  \vskip -0.1in
\end{table}
\subsection{Performance}
In this study, we employ our algorithm on the DPM++ 2m solver and UniPC solver. We employ $p_1=7$, $p_2=1.2$, $stop = 8$, $M=12$, and $N=8$ as the hyperparameters of the discrete setting in the eight-step inference when generating 1024 x 1024 and 512 x 512 images via DPM++ 2m. And its $t_{aug}$ is -1 and 8 for 512x512 and 1024x1024 resolutions, respectively. Additionally,  $t_{aug} = 6$, $p_1=12$, $p_2=1.2$, $stop = 10$, $M=12$, and $N=8$ are applied on the custom UniPC solver in the eight-step inference when generating 1024 x 1024 images. And for the six-step inference for generating 1024 x 1024-resolution images, the hyperparameters of the UniPC solver are $t_{aug}=4$, $p_1=12$, $p_2=1.2$, $ stop=10$, $M=12$, and $N=6$.  Meanwhile, for 512 x 512 images, the hyperparameters of the 8-step UniPC solver are $t_{aug}=4$, $p_1=7$, $p_2=1.2$, $stop = 9$, $M=12$, and $N=8$.  And for the customized 5-step UniPC solver for 512 x 512 images, the hyperparameters are $t_{aug}=4$, $p_1=5$, $p_2=1.2$, $stop = 6$, $M=8$, and $N=5$.  Then, the 6-step UniPC solver of the 512 x 512-resolution images has hyperparameters $t_{aug}=4$, $p_1=5$, $p_2=1.2$, $stop = 6$, $M=8$, and $N=6$.


We report the results of the inference that generate images via 10000 captions in the COCO 2014, 5000 captions in COCO 2017, and 10000 captions in the LAION dataset in \cref{tab:FIDPerformance512}, \cref{tab:FIDPerformance1024}, \cref{tab:FIDPerformance5122017}, \cref{tab:FIDPerformance1024-2017}, \cref{tab:FIDPerformance512LAION}, and \cref{tab:FIDPerformance1024-LAION}. 
Explicitly, the UniPC solver is a 2nd-order multistep solver for eight-step inference and a 1st-order multistep solver for five and six-step inference. The inference model that uses our algorithm for synthesizing 512 x 512 images and 1024 x 1024 images is Stable Diffusion v1.5 and Dreamshaper XL, respectively. 
We also compare the FID performance and the clip score of AMED-Pulging solver, progressive distillation, SDXL turbo, SDXL-Lightning, Flash DiffusionXL, Flash Diffusion, and DMD2 (\citet{yin2024improveddmd,chadebec2025flash,lin2024sdxllightning, sauer2024adversarial}).  Our approach outperforms state-of-the-art models for most of the datasets. However, some of the models (\citet{yin2024improveddmd,chadebec2025flash,lin2024sdxllightning}) are trained in LAION, which do not separate the validation and the training sets, and may affect the comparison. The $w$ in the \cref{tab:FIDPerformance512} and \cref{tab:FIDPerformance1024} is the guidance scale of inference, and the number in the parentheses is the total number of function evaluations (NFEs) of the model.  The $b_1$, $b_2$, $s_1$, and $s_2$ in the Free-U decorator is 1.1, 1.1, 0.9, 0.2.
\begin{table}[htbp]
  \caption{Results of 512 x 512 image generation.  COCO 2014}
  \label{tab:FIDPerformance512}
    \begin{center}
    \begin{small}
      \begin{sc}
  {\large
    \resizebox{\columnwidth}{!}{
  \begin{tabular}{@{}lcccccc@{}}
    \toprule
    Solver & steps & scale & use Free-U & apply stage & FID & CLIP-Score \\
    \midrule
    DPM++ 2m & 8 & 7.5 & No & Never & 15.92 & 31.32\\
    DPM++ 2m & 8 & 5.5 & No & Never & 15.7 & 31.07\\
    UniPC & 5 & 7.5 & Yes & 4-th step & 19.35 & 30.63\\
    UniPC & 6 & 7.5 & Yes & 4-th step & 16.41 & 30.97\\
    UniPC & 8 & 7.5 & Yes & 4-th step & 16.00 & 31.15\\
    UniPC & 5 & 5.5 & Yes & 4-th step & 15.19 & 30.93\\
    \textbf{UniPC} & \textbf{6} & \textbf{5.5} & \textbf{Yes} & \textbf{4-th step} & \textbf{14.72} & \textbf{31.08}\\
    UniPC & 8 & 5.5 & Yes & 4-th step & 15.26 & 31.12\\
    DPM++ 2m & 20 & 7.5 & No & Never & 17.3 & 30.97\\
    DPM++ 2m & 20 & 5.5 & No & Never & 18.68 & 30.97\\
    Flash Diffusion & 5 & None & No & Never & 17.34 & 30.45\\
    Flash Diffusion & 6 & None & No & Never & 17.94 & 30.59\\
    Flash Diffusion & 8 & None & No & Never & 19.03 & 30.37\\
    AMED Pulgin & 8 & 7.5 & No & Never & 19.07 & 31.15\\
    SDXL turbo & 4 & None & No & Never & 23.24 & 31.66\\
    \bottomrule
  \end{tabular}
  }
  }
  \end{sc}
    \end{small}
  \end{center}
  \vskip -0.1in
\end{table}
\begin{table}[htbp]
  \caption{Results of 512 x 512 image generation.  COCO 2017}
  \label{tab:FIDPerformance5122017}
    \begin{center}
    \begin{small}
      \begin{sc}
  {\large
    \resizebox{\columnwidth}{!}{
  \begin{tabular}{@{}lcccccc@{}}
    \toprule
    Solver & steps & scale & use Free-U & apply stage & FID & CLIP-Score \\
    \midrule
    DPM++ 2m & 8 & 7.5 & No & Never & 22.40 & 31.23\\
    DPM++ 2m & 8 & 5.5 & No & Never & 22.35 & 31.35\\
    UniPC & 5 & 7.5 & Yes & 4-th step & 22.70 & 30.99\\
    UniPC & 6 & 7.5 & Yes & 4-th step & 24.02 & 31.04\\
    UniPC & 8 & 7.5 & Yes & 4-th step & 24.55 & 30.99\\
    UniPC & 5 & 5.5 & Yes & 4-th step & 21.31 & 30.92\\
    \textbf{UniPC} & \textbf{6} & \textbf{5.5} & \textbf{Yes} & \textbf{4-th step} & \textbf{20.82} & \textbf{31.04}\\
    UniPC & 8 & 5.5 & Yes & 4-th step & 21.71 & 31.07\\
    DPM++ 2m & 20 & 7.5 & No & Never & 25.97 & 31.21\\
    DPM++ 2m & 20 & 5.5 & No & Never & 23.75 & 31.10\\
    Flash Diffusion & 5 & None & No & Never & 23.56 & 30.64\\
    Flash Diffusion & 6 & None & No & Never & 24.16 & 30.60\\
    Flash Diffusion & 8 & None & No & Never & 25.35 & 30.55\\
    AMED Pulgin & 8 & 7.5 & No & Never & 25.50 & 31.17\\
    SDXL turbo & 4 & None & No & Never & 30.92 & 31.66\\
    \bottomrule
  \end{tabular}
  }
  }
  \end{sc}
    \end{small}
  \end{center}
  \vskip -0.1in
\end{table}
\begin{table}[htbp]
  \caption{Results of 512 x 512 image generation. Laion}
  \label{tab:FIDPerformance512LAION}
    \begin{center}
    \begin{small}
      \begin{sc}
  {\large
    \resizebox{\columnwidth}{!}{
  \begin{tabular}{@{}lcccccc@{}}
    \toprule
    Solver & steps & scale & use Free-U & apply stage & FID & CLIP-Score \\
    \midrule
    DPM++ 2m & 8 & 7.5 & No & Never & 17.74 & 32.42\\
    DPM++ 2m & 8 & 5.5 & No & Never & 17.52 & 32.41\\
    UniPC & 5 & 7.5 & Yes & 4-th step & 20.43 & 30.87\\
    UniPC & 6 & 7.5 & Yes & 4-th step & 17.56 & 31.63\\
    UniPC & 8 & 7.5 & Yes & 4-th step & 17.38 & 32.08\\
    UniPC & 5 & 5.5 & Yes & 4-th step & 17.44 & 31.73\\
    \textbf{UniPC} & \textbf{6} & \textbf{5.5} & \textbf{Yes} & \textbf{4-th step} & \textbf{16.84} & \textbf{32.08}\\
    UniPC & 8 & 5.5 & Yes & 4-th step & 17.32 & 32.26\\
    DPM++ 2m & 20 & 7.5 & No & Never & 18.30 & 32.67\\
    DPM++ 2m & 20 & 5.5 & No & Never & 17.33 & 32.47\\
    Flash Diffusion & 5 & None & No & Never & 15.40 & 31.60\\
    Flash Diffusion & 6 & None & No & Never & 15.80 & 31.63\\
    Flash Diffusion & 8 & None & No & Never & 16.87 & 31.57\\
    AMED Pulgin & 8 & 7.5 & No & Never & 18.06 & 32.44\\
    SDXL turbo & 4 & None & No & Never & 22.25 & 33.02\\
    \bottomrule
  \end{tabular}
  }
  }
  \end{sc}
    \end{small}
  \end{center}
  \vskip -0.1in
\end{table}
\begin{table}[htbp]
  \caption{Results of 1024 x 1024 image generation.  COCO 2014}
  \label{tab:FIDPerformance1024}
    \begin{center}
    \begin{small}
      \begin{sc}
  {\large
    \resizebox{\columnwidth}{!}{
  \begin{tabular}{@{}lcccccc@{}}
    \toprule
    Solver & steps & scale & use Free-U & apply stage & FID & CLIP-Score \\
    \midrule
    \textbf{DPM++ 2m} & \textbf{8} & \textbf{7.5} & \textbf{No} & \textbf{Never} & \textbf{17.84} & \textbf{31.62}\\
    DPM++ 2m & 8 & 5.5 & No & Never & 19.62 & 31.58\\
    UniPC & 8 & 7.5 & Yes & 4-th step & 18.18 & 31.23\\
    UniPC & 8 & 5.5 & Yes & 4-th step & 17.87 & 31.40\\
    UniPC & 6 & 7.5 & Yes & 4-th step & 18.80 & 31.42\\
    UniPC & 6 & 5.5 & Yes & 4-th step & 19.07 & 31.45\\
    Flash Diffusion & 6 & None & No & Never & 22.23 & 31.00\\
    Flash Diffusion & 8 & None & No & Never & 23.25 & 30.89\\
    SDXL-Lightning & 8 & None & No & Never & 21.07 & 31.11\\
    DMD2 & 4 & None & No & Never & 19.64 & 31.64\\
    \bottomrule
  \end{tabular}
  }
  }
  \end{sc}
    \end{small}
  \end{center}
  \vskip -0.1in
\end{table}
\begin{table}[htbp]
  \caption{Results of 1024 x 1024 image generation.  COCO 2017}
  \label{tab:FIDPerformance1024-2017}
    \begin{center}
    \begin{small}
      \begin{sc}
  {\large
    \resizebox{\columnwidth}{!}{
  \begin{tabular}{@{}lcccccc@{}}
    \toprule
    Solver & steps & scale & use Free-U & apply stage & FID & CLIP-Score \\
    \midrule
    DPM++ 2m & 8 & 7.5 & No & Never & 24.42 & 31.71\\
    DPM++ 2m & 8 & 5.5 & No & Never & 26.04 & 31.56\\
    UniPC & 8 & 7.5 & Yes & 4-th step & 24.58 & 31.21\\
    \textbf{UniPC} & \textbf{8} & \textbf{5.5} & \textbf{Yes} & \textbf{4-th step} & \textbf{24.38} & \textbf{31.38}\\
    UniPC & 6 & 7.5 & Yes & 4-th step & 25.29 & 31.44\\
    UniPC & 6 & 5.5 & Yes & 4-th step & 25.20 & 31.42\\
    Flash Diffusion & 6 & None & No & Never & 28.74 & 30.96\\
    Flash Diffusion & 8 & None & No & Never & 29.63 & 30.88\\
    SDXL-Lightning & 8 & None & No & Never & 27.89 & 31.08\\
    DMD2 & 4 & None & No & Never & 26.19 & 31.65\\
    \bottomrule
  \end{tabular}
  }
  }
  \end{sc}
    \end{small}
  \end{center}
  \vskip -0.1in
\end{table}
    
    
\begin{table}[t]
  \caption{Results of 1024 x 1024 image generation.Laion}
  \label{tab:FIDPerformance1024-LAION}
    \begin{center}
    \begin{small}
      \begin{sc}
  {\large
    \resizebox{\columnwidth}{!}{
  \begin{tabular}{@{}lcccccc@{}}
    \toprule
    Solver & steps & scale & use Free-U & apply stage & FID & CLIP-Score \\
    \midrule
    DPM++ 2m & 8 & 7.5 & No & Never & 19.25 & 32.22\\
    DPM++ 2m & 8 & 5.5 & No & Never & 20.82 & 32.07\\
    UniPC & 8 & 7.5 & Yes & 4-th step & 21.59 & 31.73\\
    UniPC & 8 & 5.5 & Yes & 4-th step & 20.41 & 32.06\\
    UniPC & 6 & 7.5 & Yes & 4-th step & 22.51 & 31.82\\
    UniPC & 6 & 5.5 & Yes & 4-th step & 22.33 & 31.87\\
    Flash Diffusion & 6 & None & No & Never & 20.88 & 31.64\\
    Flash Diffusion & 8 & None & No & Never & 21.30 & 31.61\\
    SDXL-Lightning & 8 & None & No & Never & 20.45 & 32.50\\
    \textbf{DMD2} & \textbf{4} & \textbf{None} & \textbf{No} & \textbf{Never} & \textbf{16.51} & \textbf{33.28}\\
    \bottomrule
  \end{tabular}
  }
  }
  \end{sc}
    \end{small}
  \end{center}
  \vskip -0.1in
\end{table}
    
We also calculate the PRD in a 6-step inference by sampling 10k captions from COCO 2014 and getting the corresponding image in the COCO dataset as a reference.  
Our experiment shows that the generative diversity of the flash diffusion in the 6-step is less than that of us.  The guidance scale of our sampler is 5.
\begin{figure}[htbp]
  \centering
  \includegraphics[width=\columnwidth]{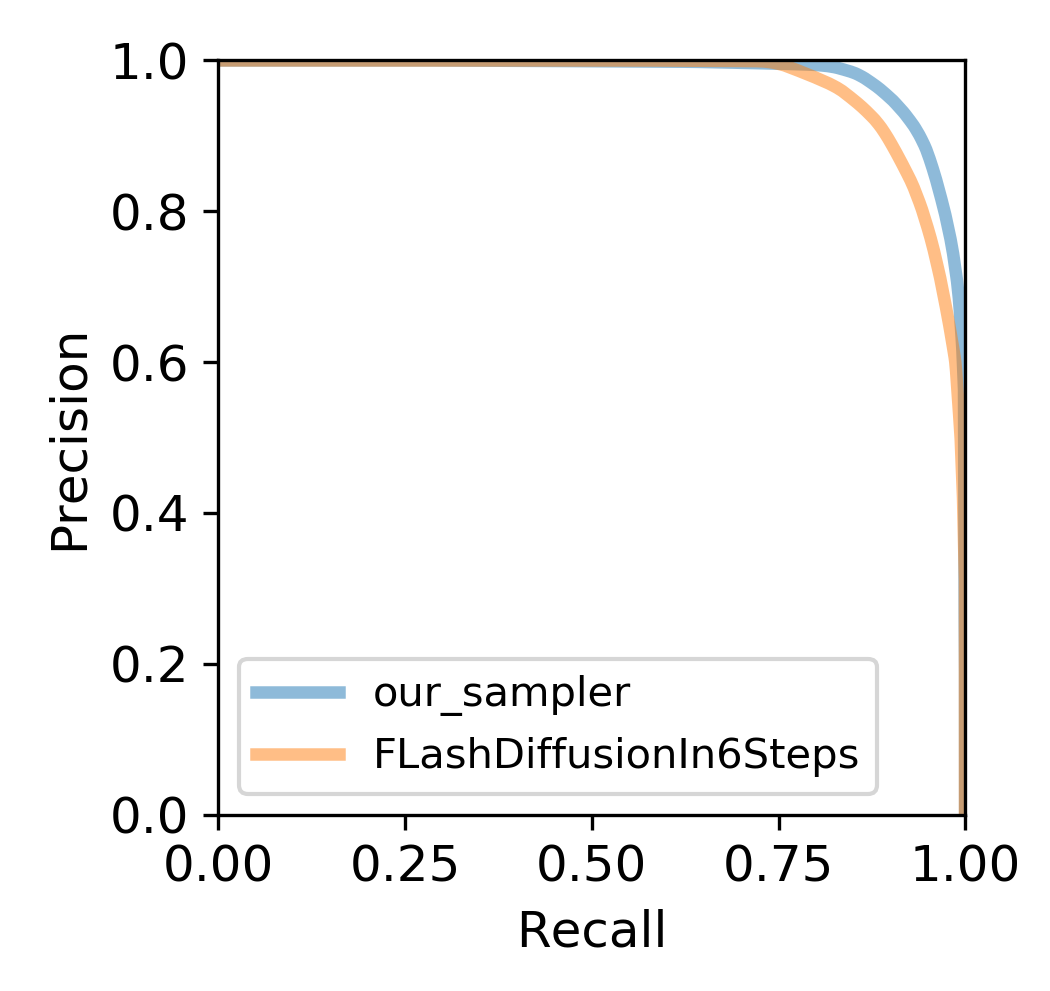}
  \caption{The comparison of PRD between our 6-step inference sampler and the FLash diffusion.}
  \label{compareDifferentTimeScheduleInPRD}
\end{figure}
\section{Conclusion}
In this study, we propose a plug-in method to enhance the sampling algorithm. By fully harnessing the potential capacity of the latent diffusion model---such as modifying the skip-connection at the right time, using the property of the decoder, ignoring a small amount of noise in the final feature to reschedule the discretizing method, and choosing the proper ODE solver via truncation error analysis---we can sample a high-quality image with limited inference budget and avoid the need for additional training.

\subsection{Limitation and the future work}
In future research, we will explore whether it is possible to find a better scheduler in a low-step inference process via an automated algorithm, and whether this new time scheduler is available to different latent diffusion models. 

\nocite{langley00}

\bibliography{example_paper}
\bibliographystyle{icml2026}

\newpage
\appendix
\onecolumn
\section{Proof of \cref{thm:loss}}

Firstly, consider the well-known probably approximately correct (PAC)\cite{hafez2020sample} deduction from Hoeffding's inequality, we have: 

\begin{equation}
   \Delta(s) = E_{X, Y}[\mathcal{L}(f^s(X) ,Y)] - \frac{1}{m} \Sigma_{i=1}^m\mathcal{L}(f^s(x_i),y_i)\leq\sqrt{\frac{\log{|\mathcal{H}|+\log{\frac{2}{\sigma}}}}{2o}},\label{eq:HoeffdingInequality}
\end{equation}
with probability approaching 1, 
where $o$ is a given constant for each hypothesis space

And input $X^{(m)}=\{x_1,x_2,...,x_m\}$, in which sample $x_i$ samples i.i.d from $\mathcal{X}$, is a sample from the $\mathcal{X}^{(m)}$.  
The $|\mathcal{H}|\leq2^{|\mathcal{X}^{(m)}|}$.
Then, for the typical set  $\mathcal{A}_{\epsilon}^{(m)}$ of the $\mathcal{X}^{(m)}$, that the probability of $X^{(m)}$ in $\mathcal{A}_{\epsilon}^{(m)}$ is bigger than $1-\epsilon$, and its cardinality $|\mathcal{A}_{\epsilon}^{(m)}|\leq 2^{m(H(x)+\epsilon)}$, then, when the batch size itself is large and the $\epsilon$ is small enough, we have:
\begin{equation}
   \Delta(s)\leq\sqrt{\frac{ 2^{mH(X)}+\log{\frac{2}{\sigma}}}{2o}}
  \label{eq:EntropyInequality}
\end{equation}
Where $H(X)$ is the entropy of the $X$.
Then, seperate the neural network into two parts: decoder $g^s$ and encoder $\phi^s$, and $f^s=g^s \circ \phi^s$, where we use $\circ$ to denote the composition of functions.  And we use $h^s_i$ to denote the $i$-th layer of the $n$-layer neural network. Explicilty, $\phi^s_l = h^s_l \circ h^s_{l-1}\circ \dots \circ h^s_1$ and $g^s_l = h^s_n\circ h^s_{n-1}\circ\dots\circ h^s_{l+1}$.
Then, the cardinality of the hypothesis space of the decoder $g^s_l$ is constrained by the number of input features $z_l^s$. And for the hypothesis space $\mathcal{H}_g$ with input $Z_l^s$:
\begin{equation}
   \Delta(s)\leq\sqrt{\frac{ 2^{mH(Z_l^s)}+\log{\frac{2}{\sigma}}}{2o}}
  \label{eq:EntropyInequality}
\end{equation}
And then we consider the conditional typical set given by the $p(Z_l^s|Y)$.  When the $Y$ is given, the cardinality of the typical set of $\{{Z_l^s}^{(1)},{Z_l^s}^{(1)},\dots,{Z_l^s}^{(m)}\}$ is $2^{mH(Z_l^s|Y)}$.
Hence for each decoder $g^s_l$ with input $Z_l^s$ that having a mutual information constrain $I(Z_l^s;Y)$, the total possible middle feature set is equal to:
\begin{equation}
   |\mathcal{H}_g| = \frac{2^{H(Z_l^s)}}{2^{H(Z_l^s|Y)}}=2^{I(Z_l^s;Y)}
  \label{eq:mutalInformationInequality}
\end{equation}

Then, for each decoder $g^s_l$, we have
\begin{equation}
   \Delta(s_g)\leq\sqrt{\frac{2^{mI(Y;Z_l^s)}+log\frac{2}{\sigma}}{2o}}
  \label{eq:IBBound}
\end{equation}
Meanwhile, for each encoder $\phi_l^s$, we use $I(\phi_l^s;S)$ to constrain the overfitting.  Explicitly, the cardinality of the hypothesis space $\mathcal{H_\phi}$ of the encoder is $2^{I(\phi_l^s; S)}$ \cite{kawaguchi2023does}. 
And then, we mulitiply the cardinality of two hypothesis spaces $\mathcal{H}_g$ and $\mathcal{H}_\phi$, to get the hypothesis space of $f_l^s$ whose the decoder is constrained with mutual information, and the encoder is in the $\mathcal{H_\phi}$, we have:
\begin{equation}
   \Delta(s_l)\leq\sqrt{\frac{2^{mI(Y;Z_l^s)}+{I(\phi_l^s; S)}+log\frac{2}{\sigma}}{2o}}
  \label{eq:encoderBound}
\end{equation}

And the $\Delta(s)$, when each encoder is constrained by a mutual information $I(\phi_l^s;S)$ to prevent overfiting, is:
\begin{equation}
   \Delta(s)\leq \underset{l}{\arg\min} \sqrt{\frac{2^{mI(Y;Z_l^s)}+{I(\phi_l^s; S)}+log\frac{2}{\sigma}}{2o}}
  \label{eq:RealIBBound}
\end{equation}
Then, if we ignore the order of output in a batch of validation data, we have 
\begin{equation}
   \Delta(s)\leq \underset{l}{\arg\min} \sqrt{\frac{2^{mI(Y;Z_l^s)}+{I(\phi_l^s; S)}+log\frac{2}{\sigma}}{2o* A_m^m}}
  \label{eq:RealIBBoundFull}
\end{equation}
and A is the arrangement number, which means the mutual information gives a bound on the generalization error. \hfill   $\Box$

\section{Proof of Proposition \ref{prop:collapse}}
For the generalization error $E_{X,Y}[{\mathcal{L}(f^s(X),Y)}]$, we have
\begin{align}
      \mathcal{L}(f^s(x),y)&=(f^s(x)-y)^2\\ 
    &= (f^s(x)-E[y|x]+E[y|x]-y)^2\\
    &= (f^s(x)-E[y|x])^2+(E[y|x]-y)^2+2(f^s(x)-E[y|x])(E[y|x]-y). \label{eq:loss}
\end{align}
Recall that 
\begin{align}
     E_{X,Y}[\mathcal{L}]&=\int_x\int_y\mathcal{L}(f^s(x),y)p(x,y)dxdy.\label{eq:Eloss}
\end{align}
By \eqref{eq:Eloss}, we compute first and third term on the right hand side of \eqref{eq:loss}:
\begin{align}
   \int_x\int_y(f^s(x)-E[y|x])^2p(x,y)dxdy &= \int_x(f^s(x)-E[y|x])^2p(x)dx,\\
   \int_x\int_y2(f^s(x)-E[y|x])(E[y|x]-y)dxdy &= 0.\\
\end{align}
Therefore, we have 
\begin{align}
       E_{X,Y}[\mathcal{L}]&=\int_x(f^s(x)-E[y|x])^2p(x)dx+ \int_x\int_y(E[y|x]-y)^2p(x,y)dxdy, \label{eq:trade1}
\end{align}
where 
\begin{align}
     \int_x(f^s(x)-E[y|x])^2p(x)dx&=E_X[(f^s(X)-\hat{f}(X))^2]+E_{X,Y}[(\hat{f}(X)-E[y|x])^2]
  \label{eq:bias-variant-trade-off-1}
\end{align}
where $E_X[(f^s(X)-\hat{f}(X))^2]$ denote the distance of the output from a particular training neural network $f^s$ and the average output $\hat{f}$. 
Note that 
\begin{align}
     \int_x\int_y(E[y|x]-y)^2p(x,y)dxdy + E_{X,Y}[(\hat{f}(X)-E[y|x])^2] = E_{X,Y}[(\hat{f}(X)-y)^2].\label{eq:trade2}
\end{align}
Combining \eqref{eq:trade1}, \eqref{eq:bias-variant-trade-off-1} and \eqref{eq:trade2}, we derive the final result. \hfill $\Box$

\end{document}